\documentclass[reprint,amsmath,amssymb,superscriptaddress]{revtex4-2}

\usepackage{graphicx}
\usepackage{dcolumn}
\usepackage{bm}
\usepackage{braket}
\usepackage{color}
\usepackage{pstricks,pst-grad,color}
\usepackage{appendix}
\usepackage{here}
\usepackage{url}
\usepackage{hyperref}
\usepackage{comment}
\usepackage{mathtools}
\usepackage[linesnumbered,lined,ruled]{algorithm2e}
\usepackage{wrapfig}

\newcommand{\mr}[1]{\mathrm{#1}}
\newcommand{\tr}[1]{\mathrm{Tr}\Bigl[#1\Bigr]}

\usepackage{tikz}
\usetikzlibrary{shapes,calc}
\usetikzlibrary{shapes.geometric,arrows.meta,decorations.markings}

\begin{document}

\title{Alpha Zero for Physics: Application of Symbolic Regression with Alpha Zero to find the analytical methods in physics}
\author{Yoshihiro Michishita}
\email{yoshihiro.michishita@a.riken.jp}
 \affiliation{RIKEN Center for Emergent Matter Science (CEMS), Wako, Saitama 351-0198, Japan}

\date{\today}

\begin{abstract}
Machine learning with neural networks is now becoming a more and more powerful tool for various tasks, such as natural language processing, image recognition, winning the game, and even for the issues of physics. 
Although there are many studies on the application of machine learning to numerical calculation and assistance of experiments, the methods of applying machine learning to find the analytical method are poorly studied.
In this paper, we propose the frameworks for developing analytical methods in physics by using the symbolic regression with the Alpha Zero algorithm, that is, Alpha Zero for physics (AZfP). As a demonstration, we show that AZfP can derive the high-frequency expansion in the Floquet systems. AZfP may have the possibility of developing a new theoretical framework in physics.  
\end{abstract}

\maketitle

\section{Introduction}
 Recently, machine learning with neural networks has shown the immensely high performance in various areas, such as the natural language processing\cite{bert, RoBERTa, GPT, llama1, llama2}, image recognition\cite{imagenet, resnet, unet, transunet} and generation\cite{greview, normalizingflow, diffusive}, mastering games\cite{AlphaGo, AlphaGoZero, AlphaZero, MuZero, r2d2}.
 Also in the area of physics, the intensive studies have begun for the application of machine learning techniques to physics problems, those are the detection of the phase transition\cite{PD1, PD2, PD3, PD4, PD5, PD6}, calculation of equilibrium states\cite{RBM1, RBM2, CarleoDNN, Ashish} or steady states\cite{RBM_OQS}, and materials informatics\cite{MI1, MI2, MI3, MI4, MI5}. In these previous studies, the machine learning techniques are used for the support of experiments or the numerical calculations. On the other hands, one may comes to a question:
 
 {\it{Can we use the machine learning techniques for developing the theoretical analysis methods?}}

Before considering how to use, let us consider what are the theoretical analysis methods. One of the main keywords may be the scale separation and the reduction, which are used in almost all areas. For examples, in nonlinear systems, we can integrate out the fast variables and perturbatively derive the effective model only with the slow variables when there is the relaxation time-scale separation among the variables.  However, in general, it is a non-trivial problem to find and separate the fast process and the slow process of the system, and we have to perform an appropriate transformation or projection and get the frame in which the scale separation is apparent. 

In this paper, we focus on utilizing reinforcement learning to find the symbolic formulation of the appropriate transformation. Recently, the symbolic physics learner is proposed by Ref.\cite{SPL}, in which they utilize the Monte-Carlo tree search\cite{MCTS} to perform the symbolic regression of the double pendulum's dynamics, and this method overcomes the conventional methods such as the genetic algorithms\cite{GP1, GP2} or the deep symbolic regression\cite{DSR}.

Inspired by this work, we propose the framework of the application of Alpha Zero\cite{AlphaZero} to the symbolic regression of the appropriate transformation, which we call ``Alpha Zero for Physics"(AZfP).

\

\section{Tree representation of equations and the game of its construction}
Here, we introduce three types of nodes for describing equations: function nodes, branch nodes, and variable nodes, shown in Figure~\ref{fig:nodes}.
The function nodes catch one edge and pass another edge toward the leaf. The function nodes can describe, for examples, $\exp$, $\sum_i$, $\frac{\partial}{\partial x}$, $\int dt$, and the like.
The branch nodes catch one edge and pass two edges toward the leaf. The branch nodes operations such as $+$, $-$, $\times$, and $-i[,]$.
The variable nodes are the leaf nodes, which catch one edge from the root and pass no edges. The variable nodes describe the variables or operators in the physics model we consider, such as $A$, $\psi_i$, or $\mathcal{H}$. We also include constants in the variable nodes.

\begin{figure}
    \centering
    \includegraphics[width=0.98\linewidth]{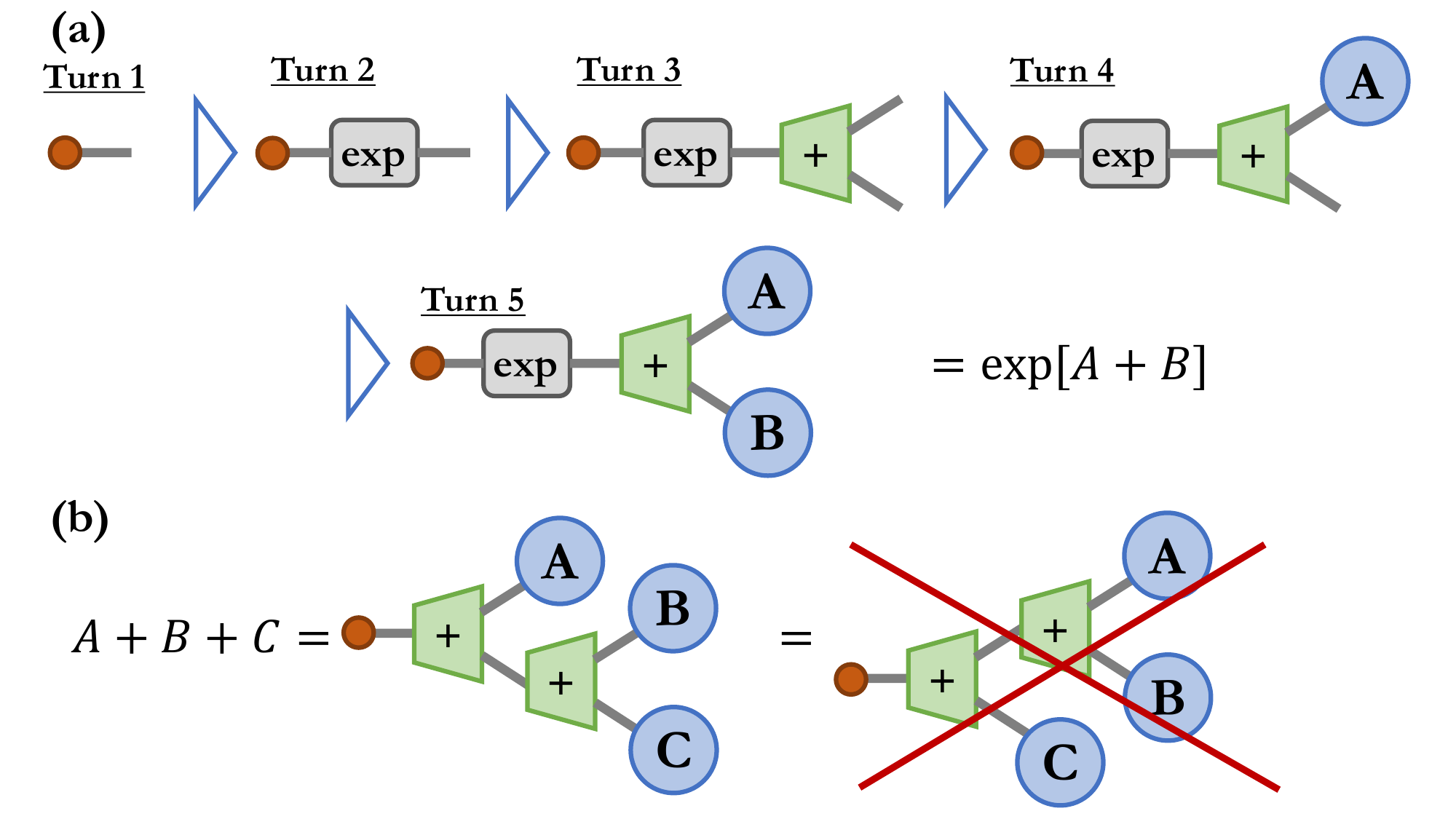}
    \caption{(a)An example of the tree construction and its procedures. (b) An example of the forbidden tree.}
    \label{fig:example}
\end{figure}

We can construct a tree graph by starting from the root edge, adding the nodes and edges, and finishing by connecting all edges to (the variable) nodes. The completed tree describes a specific equation. An example is shown in Figure~\ref{fig:example}. We can describe any equation by a tree with these three types of nodes.

Therefore, finding the equation with some desirable properties is equal to finding a series of the tree's nodes with the desirable properties from the possible nodes or finding the best strategy from the action space. If we can define the reward, which becomes large when the equation has the desirable properties, the problem is a game to find the best strategy with the highest score. We note that how to define the score depends on the kind of physics problem.

For the well-defined game, we also pose the additional rules as follows:
\begin{itemize}
\item{We must add the nodes from the top to the bottom. (Definition of the procedure of the game.)}

\item{The number of the variable nodes ($N_{\mr{var}}$) must be equal or less than the number of the branch nodes ($N_\mr{br}$) plus one, that is $N_{\mr{var}} \leq N_\mr{br}+1$. The equation is completed when the equality is satisfied, and the game ends. (Definition of the condition for the termination of the game.)}

\item{The same branch cannot be consecutive. (Due to the symmetry of the operation. For an example, $A+B+C$ must be expressed in $A+(B+C)$, not in $(A+B)+C$ as shown in Figure~\ref{fig:example}).}

\item{If the function nodes include a certain function and its inverse function, they cannot be consecutive. (Forbid of the redundancy of the game as same as ``Sennichite" in Shogi game.)}
\end{itemize}

Obeying these rules, we search for the best strategy with AZfP.

\begin{table}
  \begin{tabular}{|c||c|c|c|} 
    \hline node type & Function & Branch & Variable \\ \hline
    Graph &
    \begin{minipage}{16mm}
      \centering
      \scalebox{0.2}{\includegraphics{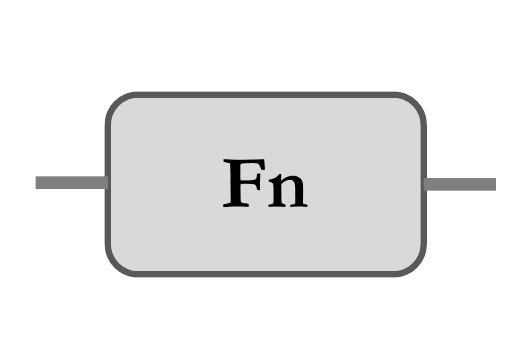}}
    \end{minipage} &
    \begin{minipage}{16mm}
      \centering
      \scalebox{0.2}{\includegraphics{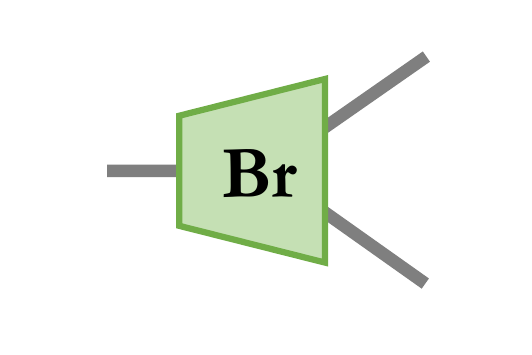}}
    \end{minipage} &
    \begin{minipage}{16mm}
      \centering
      \scalebox{0.2}{\includegraphics{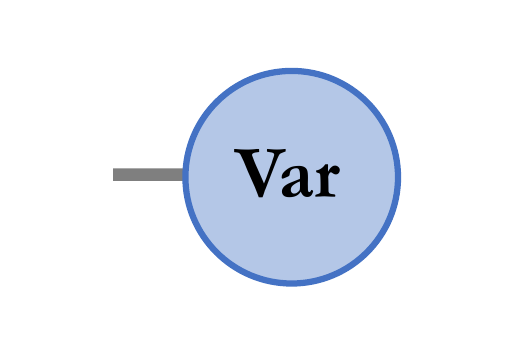}}
    \end{minipage} \\ \hline
    Examples & $\exp, \sum_i, \int dt, \frac{\partial}{\partial x}$ & $+, -, \times, -i[,], \{,\}$ & $x, \psi_i, H$ \\ \hline
  \end{tabular}
  \caption{Table of the types of the nodes for describing equations in a tree graph.}
  \label{fig:nodes} 
\end{table}

\

\section{Algorithms of Alpha Zero for Physics}
Here, we briefly introduce the algorithms of AZfP;  further details can be found in the Supplementary Materials(SM).\cite{supple} The algorithms of AZfP are based on those of original Alpha Zero\cite{AlphaZero}.
Alpha Zero is divided into the search algorithms and the architecture of the neural network. 
We save and update the statistics of the results of the search in the nodes $(s,a)$ of the policy tree, which are ${N(s,a), Q(s,a), P(s,a)}$ while searching for the best policy. We note that $s$ represents the state of the making equation, and $a$ describes the action or the next node to be added to the equation tree. $N(s,a)$ is the visit count of the nodes of policy trees in the search. $Q(s,a) = \mr{max}_{a'\in \mathcal{A}}Q(s',a')$ is the action value where  $s'$ is the next state after performing $a$ in the state $s$. $P(s,a)$ is the prior probability of selecting the action $a$ in the state $s$, and is given by the neural network. By utilizing these statistics, we perform the PUCT algorithms for search, which reads,
\begin{eqnarray}
    a_t &=& \mr{max}_{a} \Bigl(Q(s_t, a) + U(s_t, a)\Bigr),\label{ucb}\\
    U(s_t,a) &=& C(s) P(s,a) \sqrt{N(s,a)} / (1+N(s,a)),\\
    C(s) &=& \mr{log}\Bigl[ \bigl(1 + \sum_a N(s,a) + c_{\mr{b}}\bigr)/c_{\mr{b}} + c_{\mr{i}}\Bigr],
\end{eqnarray}
where $c_b$ and $c_i$ are the hyperparameters which decide the agent's personality for searching. We note that the PUCT algorithm is a kind of the MCTS algorithms. After searching and storing the statistics (we call this PUCT simulations), AZfP play the game by performing the action which is most explained in the simulation. 

In the SPL, they use the MCTS without neural networks, while they include the transplantation of the high-score trees for searching deep trees efficiently, which may be the remnant of the genetic algorithms. Although the transplantation may leads high score trees quickly, it should equal to the action space change in the simulation and may ruin the MCTS(UCB) algorithm. AZfP instead utilizes the neural networks for efficient search, without causing such situations.  
Moreover, inspired by the SPL, the agent in AZfP holds the memory of the score of some equation trees and utilizes it to guess $Q(s,a)$. If the agent does not have the memory of the score of the tree focused on, the agent evaluates the estimation value from the neural network.

\begin{figure*}\label{fig:picture}
    \centering
    \includegraphics[width=0.8\linewidth]{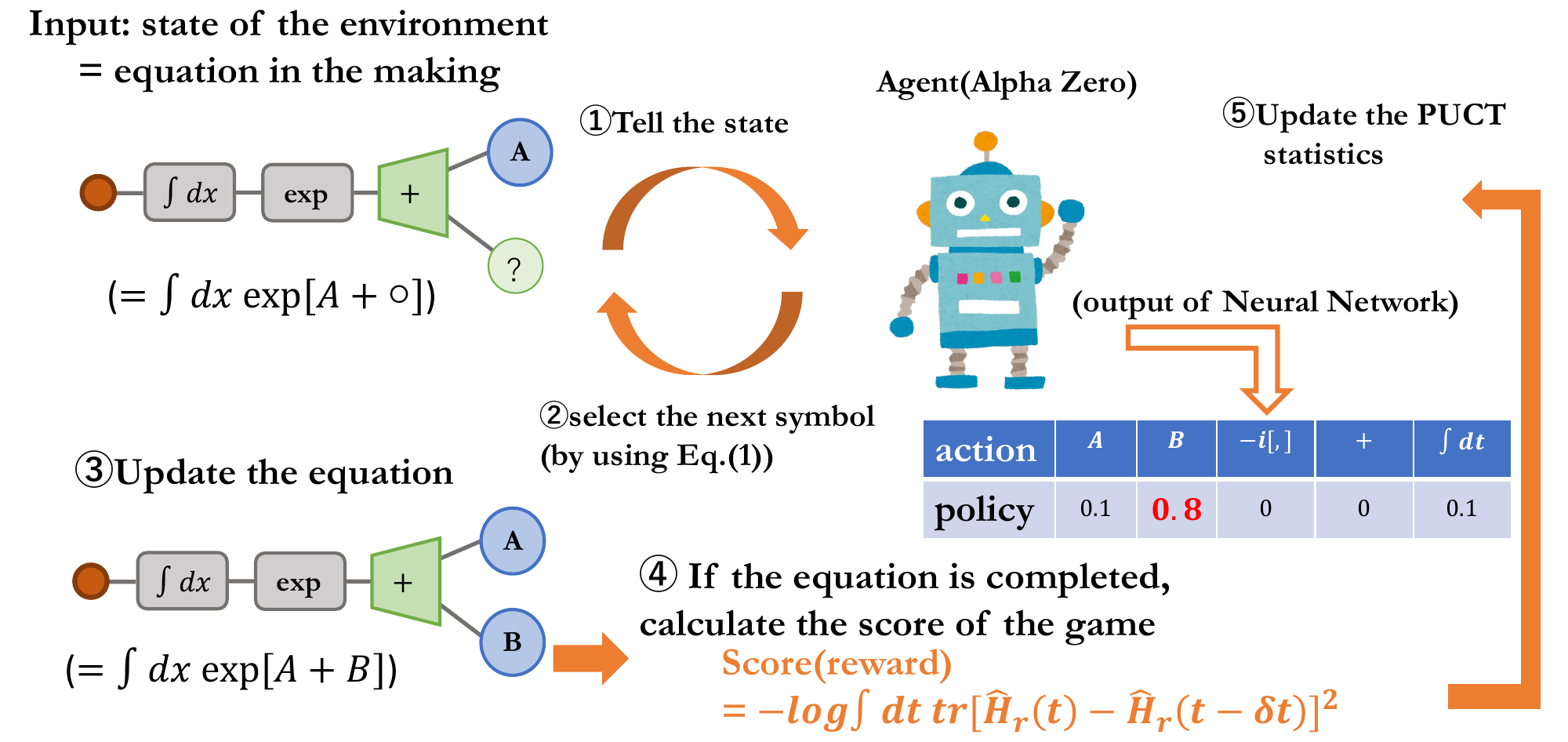}
    \caption{The picture of the five step procedure of Alpha Zero for Physics. Step1: Input the state vector to the neural network; Step2: We choose the action by Eq.(\ref{ucb}) in the simulation of PUCT search or choose the most-simulated action in the self-play; Step3: Update the state and the equation; Step4: If the equation is completed, calculate the score of the equation; Step5: Update the PUCT statistics in the simulation or train the neural network using the self-play statistics.}
\end{figure*}

\begin{figure}
    \centering
    \includegraphics[width=0.98\linewidth]{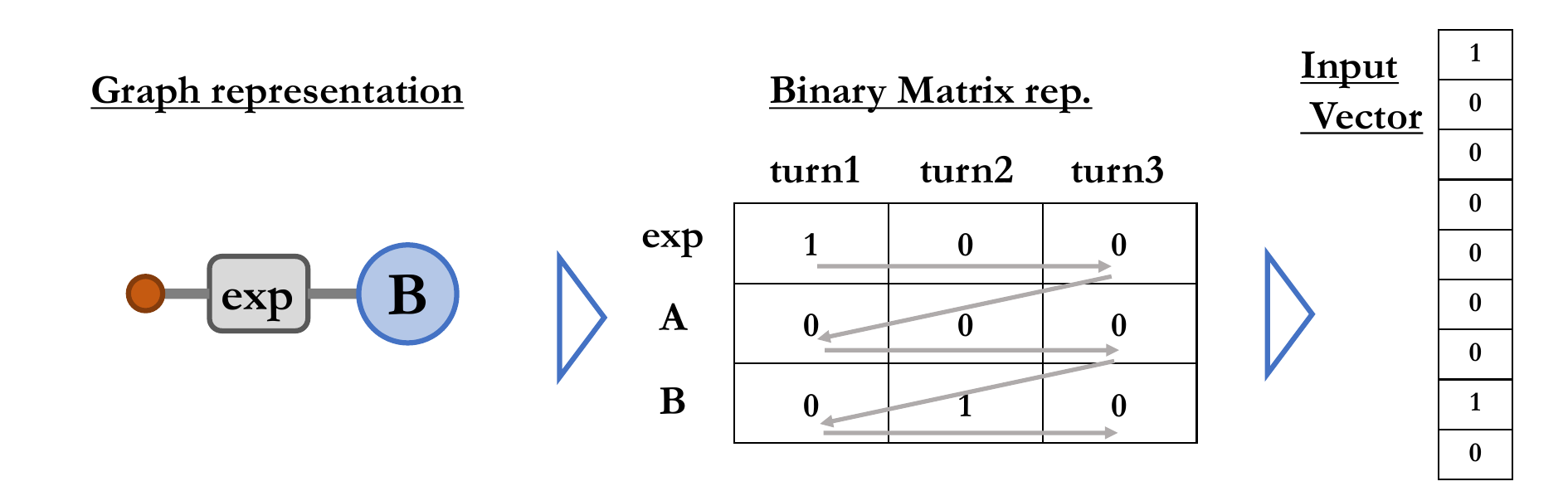}
    \caption{vector representation of equation state as an input for the case where the action space $\mathcal{A}$ consists of $\mr{exp}, A, B$ and the maximum turn is 3.}
    \label{fig:repre_st}
\end{figure}

We represent the state of the making tree for an equation in $(A*T_\mr{max})$-dimension binary vector as shown in Figure~\ref{fig:repre_st}, where $A$ is the number of kinds of action and $T_\mr{max}$ is the allowed maximum turn (tree length). We input this state vector to the neural networks and get $[[P(s_t,a_t) \ \mr{for} \ a_t \in \mathcal{A}], V(s_t)]$ as the output. By using this output and Eq.(\ref{ucb}), AZfP chooses the next node to be added to the tree.  The neural network is trained from the results which the agent has in the memory.
We describe the flow of the procedure of AZfP in  Figure.\ref{fig:picture}. We also show the detail algorithm in SM\cite{supple}.

\

\section{Application of Alpha Zero for Physics to periodically-driven systems}
Finally, we demonstrate that the AZfP can derive the theoretical analysis methods in periodically-driven systems, which is the high-frequency expansion to calculate the Floquet unitary transformation.

In periodically-driven systems, we consider the time-dependent Hamiltonian as
\begin{align}
    \hat{H}(t) = \hat{H}_0 + \hat{V}(t),
\end{align}
where $\hat{H}_0$ is the system Hamiltonian without the driving and $\hat{V}(t)$ is the driving term which depends on time. Under the time-dependent unitary transformation, the Schr\"{o}dinger equation  $i\frac{d}{dt}\ket{\psi(t)} = \hat{H}(t)\ket{\psi(t)}$ can be written as
\begin{eqnarray}
    i\frac{d}{dt}\ket{\tilde{\psi}(t)} &\equiv& i\frac{d}{dt} \hat{U}(t) \ket{\psi(t)}
    =\hat{H}_r(t) \ket{\tilde{\psi}(t)},\\
    \hat{U}(t) &=& \exp\Bigl[i \hat{K}(t)\Bigr],\\
    \hat{H}_r(t) &\equiv & \hat{U}(t) \hat{H}(t)\hat{U}^\dagger(t) + i \bigl(\frac{d}{dt}\hat{U}(t)\bigr)\hat{U}^\dagger(t)\\
    &=&\hat{U}(t)\Bigl(\hat{H}(t) - i \frac{d}{dt}\Bigr)\hat{U}^\dagger(t),\label{Hr}
\end{eqnarray}
where we call $\hat{H}_r(t)$ as the dressed Hamiltonian, which describes the dynamics of $\ket{\tilde{\psi}(t)}$ from the reference frame of $\hat{U}(t)$.

When we focus on the periodically-driven systems, where $\hat{H}(t)=\hat{H}(t+T)$ and $\hat{V}(t)=\hat{V}(t+T)$ are satisfied, it is known that there is the periodic time-dependent unitary transformation $\hat{U}_F(t) = \hat{U}_F(t+T)$, which make the dressed Hamiltonian time-independent, that is, $\hat{H}_r(t) = \hat{H}_F$. We call $\hat{H}_F$ Floquet Hamiltonian,  $\ket{\tilde{\psi}(t)} = \hat{U}_F(t)\ket{\psi(t)}$, whose dynamics is described by the Floquet Hamiltonian, the exact rotating frame (RF), and $\hat{K}(t)$ the micromotion operator or the kick operator. 
In the high-frequency regime, we can construct this exact rotating frame perturbatively with high-frequency expansion such as the Floquet-Magnus expansion\cite{PhysRevLett.116.120401, KUWAHARA201696} and van-Vleck expansion\cite{doi:10.1080/00018732.2015.1055918, Eckardt_2015}. If we calculate up to the finite order $\mathcal{O}(1/\Omega^n)$, the dressed Hamiltonian slightly depends on time by $\mathcal{O}(1/\Omega^{n+1})$.
Fortunately, when we calculate the heating rate of the system under the high-frequent driving, it is more beneficial to calculate the high-frequency expansion up to the second or third order \cite{PhysRevLett.128.050604} and we call this benefitial RF an appropriate RF.

In the following, supposing that the physicists in the world do not know the high-frequency expansion methods and the appropriate RF but want to derive the appropriate time-periodic unitary transformation in which the dressed Hamiltonian slightly depends on time, we ``derive" the high-frequency expansion and the appropriate RF with AZfP.

We set $H_0$ and $V(t)$ as the variable nodes. Because variables and the target equation $K(t)$ are Hermitian, we set $+$, $-i[,]$, $\frac{1}{2}\{ ,\}$ as the branch nodes which keep the Hermiticity. We set $\int dt$ as the function nodes, while it cannot include time-derivative $\partial/\partial t$ because it cause divergence when the driving is proportional to the step function. AZfP constructs $K(t)$ by combining these operators and variables under the boundary condition $K(t=0) = 0$, and calculate the dressed Hamiltonian $H_r(t) = U(t)H(t)U^{\dagger}(t) -iU(t) \{U(t+\delta t) - U(t)\}/\delta t$.

Because we want to get the RF in which the time-dependence of the dressed Hamiltonian is small, we define the reward of the game as,
\begin{eqnarray}
    r &=& -\rm{log}\Bigl[\sum_{n=1}^{N_\mr{t}} ||H_r(n\delta t) - H_r((n-1)\delta t)|| \delta t\Bigr],
\end{eqnarray}
where $||M|| = \tr{M^2}$ and $\mathrm{N_{t}} = T/\delta t$ is the total time step.

Under this setup, AZfP selects the next adding node, completes the equation, and searches for the best symbolic construction of Hermitian $K(t)$ which makes the time-dependence of $H_r(t)$ smallest.
\

In order to calculate the reward, we have to introduce the concrete model and demonstrate the concrete procedure. Here, AZfP derive the symbolic representation for $K(t)$ and it should be useful even when the system is different. Thus, we should use the easiest model to calculate the reward, although we have to check in what parameter regime and under what kind of symmetries the derived symbolic representation can be used. 

For simplicity, in this letter, we analyze an interacting two-spin model under the periodic driving, which reads,
\begin{eqnarray}
    \hat{H}_0 &=& \bm{h}\cdot\sum_{i=1,2} \hat{\bm{s}}_i + \sum_{\alpha=x,y,z} J_{\alpha}\hat{s}^{\alpha}_1 \hat{s}^{\alpha}_2,\label{H_0}\\
    \hat{V}(t) &=& \bm{\xi}\cdot \sin(\Omega t) \sum_{i}\hat{\bm{s}}_i,\label{Vt}
\end{eqnarray}
where $\hat{s}_i$ represents the quantum spin operator at site $i$, $\bm{h}$ represents a static magnetic fields, $J_{\alpha}$ represents an interaction between the two spins, and $\bm{\xi}$ represents an AC magnetic fields. For the simplicity, we set $\bm{h}=(0, 0, h_z)$, $\bm{J} = (J_x,0,J_z)$, and $\bm{\xi} = (\xi, 0, 0)$.

Next, we show the results by AZfP. We set the hyperparameters as $c_b=256, c_i=1.25$, which is roughly optimized by the genetic algorithm. We use Adam \cite{ADAM} as the optimization function. The detail setup for deep neural networks in AZfP is written in the SM\cite{supple}.

\begin{figure}
    \centering
    \includegraphics[width=0.98\linewidth]{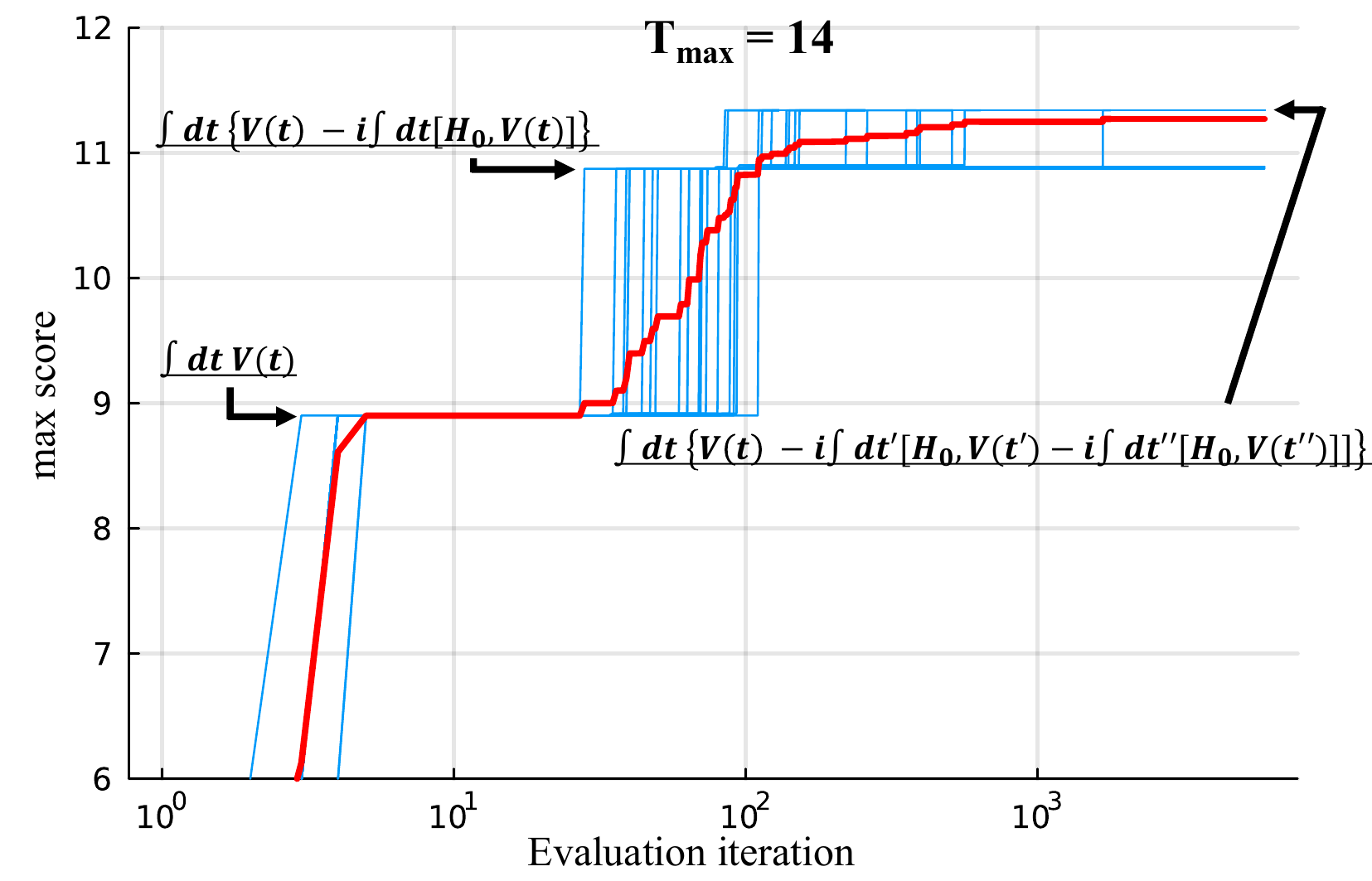}
    \caption{Searching dynamics by AZfP when we set $T_\mr{max}=14$. The y-axis represents the max score found in the search, and the x-axis represents the iteration number of calculated scores in the search. Because the search includes randomness, we have checked in 20 trials. The blue lines represent the performance of each trial, and the red line represents their average. AZfP found the symbolic representation of the unitary transformation corresponding to the first-, second-, and third-order Floque-Magnus expansion in order and also found some unphysical rotating frames with good scores. We set the model parameters as follows: $\Omega = 10, \xi = 0.4, J_z = 1.0, J_x = 0.7, h_z=0.5$.}
    \label{fig:results}
\end{figure}

Figure~\ref{fig:results} shows the best unitary transformation and its reward (score) which AZfP found when setting the maximum tree length $l_{\rm{max}} = 14$. AZfP, in order, found the first-, second-, and third-order Floquet-Magnus expansion for the rotating frame. We set the iterations of calculating the score as the x-axis because the calculation of the scores is the bottleneck for the calculation time.  Figure~\ref{fig:results} also shows that AZfP can efficiently search for the appropriate rotating frame because AZfP is able to find the third-order Floquet-Magnus expansion by almost $10^3$ times search in $T_\mr{max}=14$ case the searching space roughly consists of $4^{14} \sim 10^7$ points to search (because $|A|-N_{\mr{var}}=4$ and the variable nodes are the terminal nodes.). 
We note that other reinforcement algorithms could not find these frames efficiently as AZfP could.\footnote{We could not compare our results with the previous studies, such as the deep symbolic optimization\cite{DSR} and SPL\cite{SPL}. There are some reasons. First, their codes are not open or do not work when following their instruction in the tutorial text. Second, their codes are written for the symbolic regression of the nonlinear equation and the case in which the variable node only consists of a single variable $x$. Therefore, they seem unable to be used for our calculation.} The performance comparison with $\epsilon$-greedy algorithm and the Actor-Critic algorithm with the proximal policy optimization\cite{PPO} is written in SM\cite{supple}. 
We also note that, when we set the $T_\mr{max}=20$, AZfP also found the unphysical frame as the best frame, whose physical units of terms are not consistent. (See the SM\cite{supple}.)

\

\section{Application of Alpha Zero for Physics to other physical problem.}
Finally, we argue how the framework of AZfP can be applied to other general problems. One way is the construction of the symbolic representation effective model of unknown nonlinear dynamics such as the dynamics of active matters and weather. This direction is similar to the previous symbolic regression research. In these cases, the score is defined as the smallness of the difference between the target dynamics and the constructed equations. Another direction is the derivation of the effective model from the original Hamiltonian or nonlinear dynamics. Also, in these cases, the score should be the smallness of difference between the target dynamics and the constructed equations when the time-scale or space-scale is coarse-grained. (For example, we set the target dynamics as the five time-step by the original dynamics, and AZfP searches for the effective Hamiltonian or the nonlinear equation, which gives the target dynamics in one step.) We note that this direction has been studied recently in the context of quantum computation, in which reinforcement learning searches for the effective and possible computation in noisy quantum devices from the original computation\cite{daimon}. We note that the time for calculating the reward often becomes the bottleneck of the calculation time of AZfP, and therefore, we should define the reward to be calculated in not-so-large time.

\

\section{Conclusion and Remarks}
In this letter, we have proposed the method to construct the appropriate transformation using the Alpha Zero algorithm and demonstrate the concrete process and the results for the periodic-driven system.
Just by defining the reward as it becomes large when the tree of the equation has the desirable properties, AZfP can search for its symbolic representation. 

We remark on some of the future developments of AZfP.
In this work, we do not include the arbitrary constant as the variable nodes, assuming that the appropriate unitary transformation should be constructed only by the physical operators and their combinations and not include certain constants. When we use AZfP to derive an effective model, we should include the constant as the variable nodes and optimize constants after completing the equation tree. 

In this work, we can use $(A*T_{\mr{max}})$-dimensional vector as an input because it is not so large. We note that we should set $T_\mr{max}$ not-so-large because a too-long equation is difficult for humans to use and it may be useless. When $(A*T_{\mr{max}})$ becomes too large due to the large action space, we may have to use the recurrent architecture such as the recurrent neural networks\cite{RNN, DSR, PhySO} as the previous researches use or the RetNet\cite{RetNet}. However, we note that the recurrent architecture often makes the learning difficult\cite{DiffinRNN}.

Some previous research about symbolic regression in physics \cite{AIFeynmann, PhySO} considers the physical unit and imposes its constraint on trees. The constraint on the search space usually leads to performance improvement. The application of this technique to AZfP is left for future work.

We believe our proposed method, Alpha Zero for Physics, is applicable even in other systems and problems, and this work paves a new way to utilize machine learning to develop analytical methods in physics. 

\

\section*{Acknowledge}
YM deeply appriciates Akinori Tanaka for fruitiful discussion.
We use Julia\cite{julia} and its library Flux.jl\cite{Flux} for all calculation, and the codes we have used are uploaded in the Github\cite{github}. This work is supported by KAKENHI 23H04527 and RIKEN Special Postdoctoral Researcher Program. Computer simulations were partially done on the supercomputer of Tokyo University at the ISSP.

\bibliographystyle{apsrev4-1}
\bibliography{ML.bib}
\clearpage

\renewcommand{\thesection}{S\arabic{section}}
\renewcommand{\theequation}{S\arabic{equation}}
\renewcommand{\thefigure}{S\arabic{figure}}

\setcounter{section}{0}
\setcounter{equation}{0}
\setcounter{figure}{0}

\onecolumngrid
\begin{center}
{\large
{\bfseries Supplemental Materials for \\ ``Alpha Zero for Physics: Application of Symbolic Regression to find the analytical methods in physics'' }}
\end{center}

\vspace{10pt}

\onecolumngrid
\section{\label{app:Algorithm} Algorithm of Alpha Zero for Physics}

\begin{algorithm}[H]
\small
\SetAlgoLined
    \textbf{Input:} Action space $\mathcal{A}=(\{F\}, \{B\}, \{V\})$, reward function $R(H_r)$\;
    \textbf{Parameters:} parameters of player's properties for PUCT $\{c, c_i, c_b\}$, noise parameter in exploring $\{f, \gamma \}$, maximum turn $T_{\mr{max}}$\;
    \textbf{Output:} optimal tree $\tilde{T}^\star$ and its equivalent equation \;
    \For{\textbf{each episode}}{
        \For {\textbf{each self-play}}{
            \textbf{Initialize:} $s_0=\mr{zeros}(d), t=0, d = |A|T_{\mr{max}}$\;
            \For {\textbf{each turn for self-play}}{
                \For {\textbf{each simulation}}{
                    \While {$s_t'$ expandable and $t' < T_{\mr{max}}$}{
                        Expand child nodes and evaluate $P(s_t', a)$ with the neural networks\;
                        Choose $a_{t'+1}= \mr{argmax}_{a\in \mathcal{A}}\Bigl(Q(s_t', a) + U(s_t', a)\Bigr)$\;
                        Take action $a_{t+1}$, update $s'[(a-1)T_\mr{max}+t+1] = 1$\;
                        $s_{t'+1} \leftarrow s'$ note as visited, $t' \leftarrow t'+1$\;
                    }
                    \If{the equation tree is completed}{
                        Calculate the reward (score) and update $N(s,a)\leftarrow N(s,a)+1, Q(s,a)\leftarrow \rm{max}[Q(s,a), R(H_r)]$ by the back propagation\; 
                        save the history $(a_1, a_2, \dots, a_L)$ and its values $\{Q(s, a)\}$ in buffer\;
                    }
                }
                \textbf{Selection:} $a_t=\mr{argmax_a}[N(s_t, a_t)]$\;
                $t \leftarrow t+1$\;
            }
        }
        \For {\textbf{each training}}{
            make the image and target data $([a_1, \dots, a_t], \{Q(s_t, a_t), p(s_t, a_t)=N(s_t, a_t)/\sum_aN(s_t,a)\})$ by sampling from the buffer\;
            \textbf{Training:} train the neural networks with the loss function defined by Eq.~(\ref{app:loss})\;
        } 
    }
    \caption{Searching Algorithm with PUCT and the training of the neural networks}\label{alg:SPL}
\end{algorithm}

\section{\label{app:Architecture} Architecture of the neural networks and its learning}
As same as the Alpha Zero\cite{AlphaZero}, the neural networks of AZfP consist of a ``body"  followed by both the policy and value ``heads." The body consists of the input layer and 12 residual blocks, each of which consists of the batch-normalized layer and a single relu layer whose width is 64 with the skip connection. The width of the input layer is $A*T_\mr{max}$. The policy head consists of 3 residual blocks, and the output relu layer whose width is $A$ and the softmax function. The value head consists of 3 residual blocks and the output layer whose width is one and activation function is $\sigma(x) = 15\tanh(x/10)$.

After playing 100 games with the PUCT algorithm, we stock the equation trees $(a_1, a_2, \dots, a_L)$. We create batches of images, which is input vector converted from a tree $(a_1,\dots, a_l) \ (l\leq L)$ as in Figure~\ref{fig:repre_st}, and targets, which is $Q(s_l, a_l)$ of the node, from this stock by randomly choosing games and nodes($l\leq L$).
We define the loss function using these images and targets of the batches as follows::
\begin{eqnarray}
    l(\theta) = \frac{1}{N_\mr{batch}}\sum_{b=1}^{N_\mr{batch}} \Bigl[\bigl((v(s_{lb})-\bar{v}(s_{lb}))/10\bigr)^2 - \bm{\pi}(s_{lb})\mr{log}[\bm{p}(s_{lb})] + Cw^2 \Bigr],\label{app:loss}
\end{eqnarray}
where $v(s_{lb}) = Q(s_l, a_l)$, $\bar{v}(s_{lb})$ is the estimated value by the neural networks, $\bm{\pi}(s_{lb}) = [N(s_{lb}, a_{lb})/(\sum_a N(s_{lb}, a_{lb})) \ \mr{for} \ a_{lb}]$, $\bm{p}(s_{lb})$ is the policy output from the neural network, $C$ is the hyperparameter, and $w$ is the weight parameters of the neural networks. The first term is the mean square error for the action value, the second term is the cross entropy term for the policy, and the last term is the weight decay term. We rescale the mean square error for the action value because its norm is less than 1 in the original Alpha Zero.

\section{\label{app:Comparison} Comparison with other reinforcement learning techniques}
\begin{figure*}[h]
    \centering
    \includegraphics[width=0.48\linewidth]{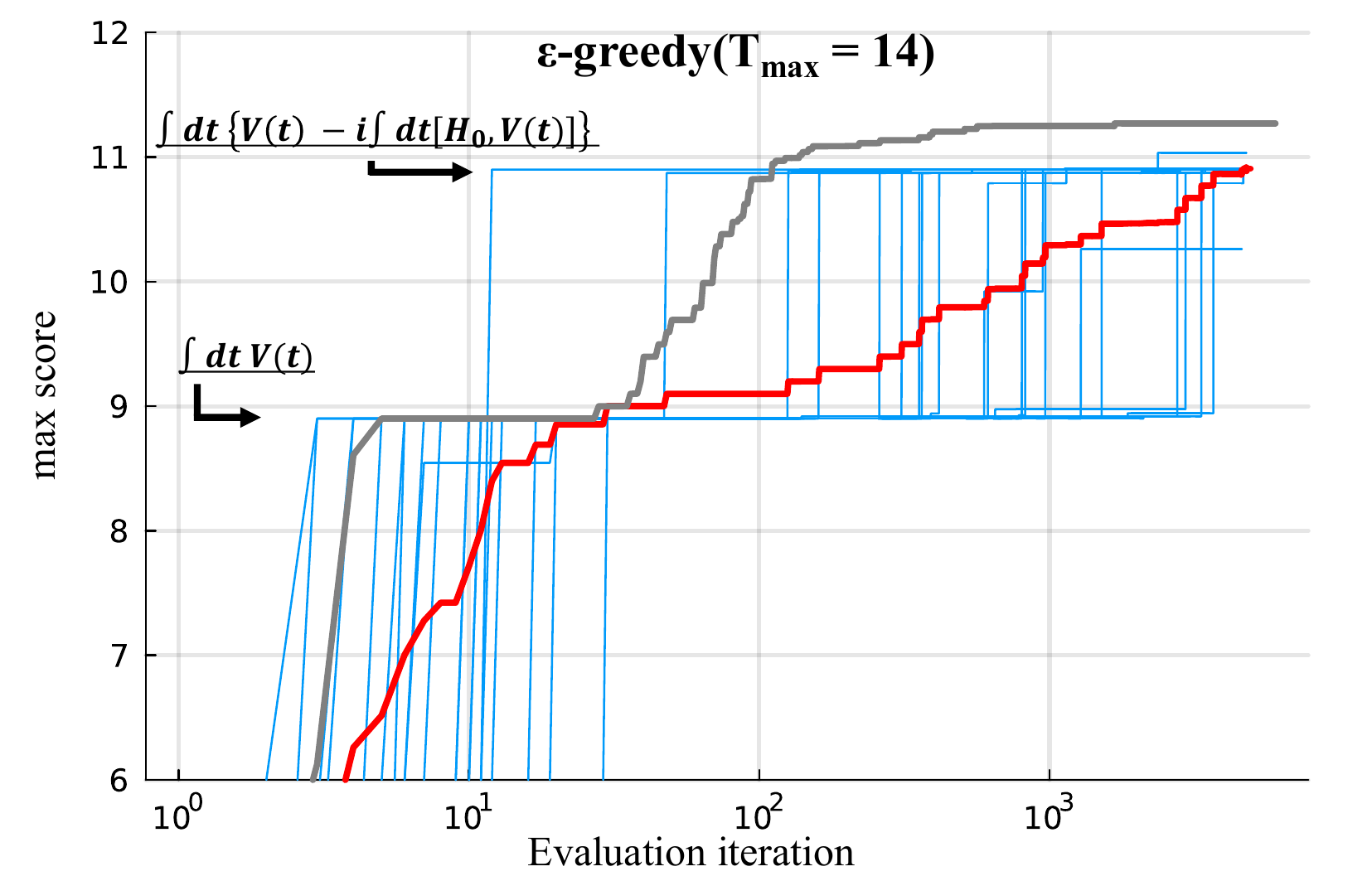}
    \includegraphics[width=0.48\linewidth]{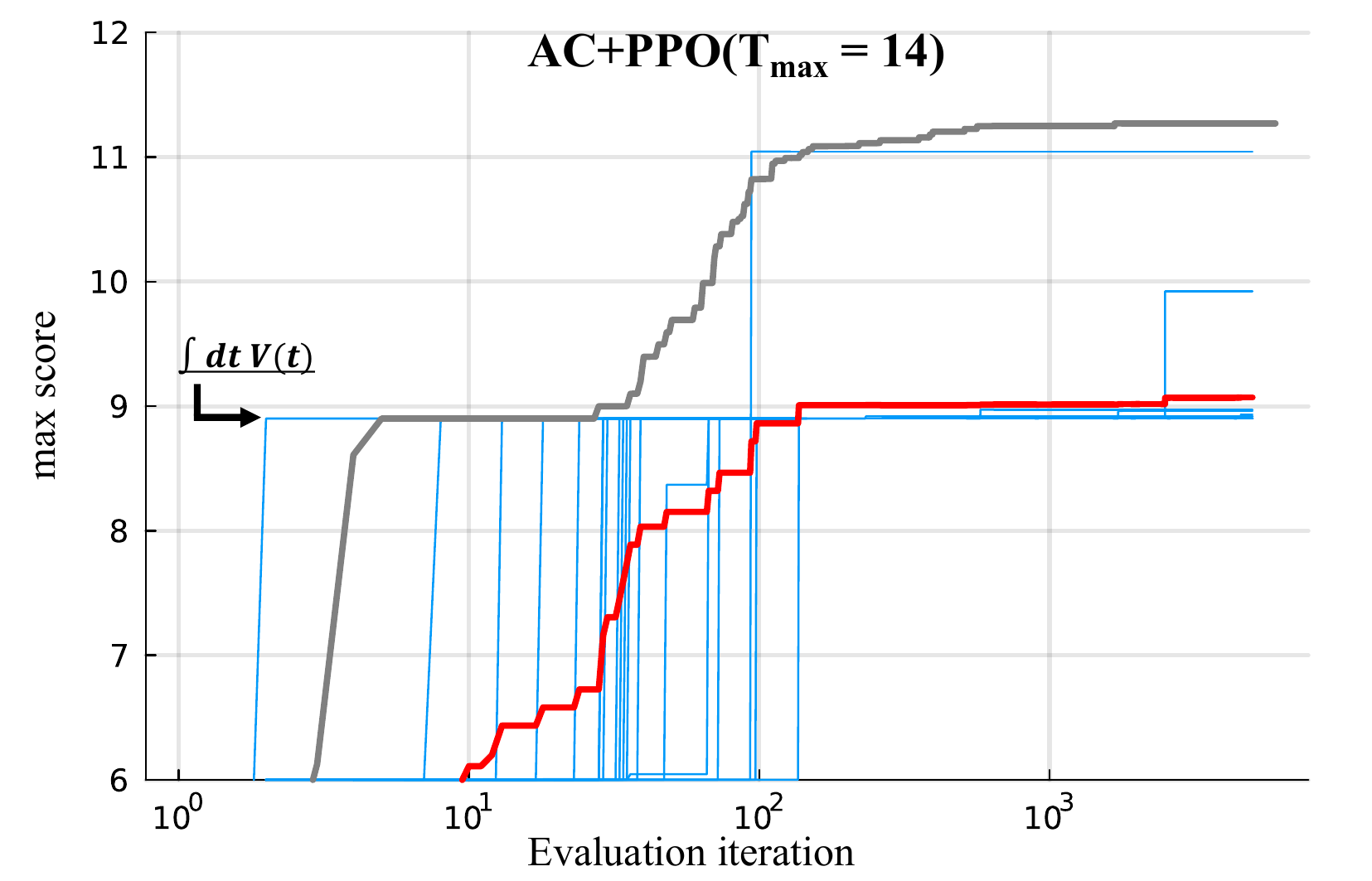}
    \caption{The scores via evaluation iteration by the $\epsilon$-greedy algorithm and the actor-critic method with PPO. We have calculated 20 trials by each methods. The blue lines show each score dynamics, the red line shows the average of them, and the grey line shows the average performance of AZfP.}
    \label{fig:compare}
\end{figure*}

Here, we show the results by the other reinforcement learning algorithms, which are the $\epsilon$-greedy algorithm and the Actor-Critic algorithm with the proximal policy optimization\cite{TRPO, PPO} when setting the maximum length is 14. Figure~\ref{fig:compare} shows that the other algorithm usually cannot find the third-order Floquet-Magnus expansion while Alpha Zero for Physics can find. The behavior of $\epsilon$-greedy algorithm should be similar to those of genetic algorithm because they explore the proximity of the state which gives the maximum reward at that time. Thus, our AZfP should show better performance than the conventional genetic algorithm for symbolic regression.
It is known that the the On-Policy algorithm such as PPO often show a bad sample efficiency. In figure.\ref{fig:compare}, AC+PPO shows the worst results and it may stem from the smallness of the samples.  

\section{\label{app:Comparison} Search for the longer equation}
Here, we show the results when we set the maximum length $T_\mr{max}=20$. When the maximum length becomes larger, the exploration space becomes larger and it becomes more difficult to find good rotating frames. In $T_\mr{max}=20$ case, the unphysical rotating frame in which the physical unit is inconsistent can have better score than the Flouqet-Magnus expansion, and our AZfP can also find them.

\begin{figure*}[h]
    \centering
    \includegraphics[width=0.48\linewidth]{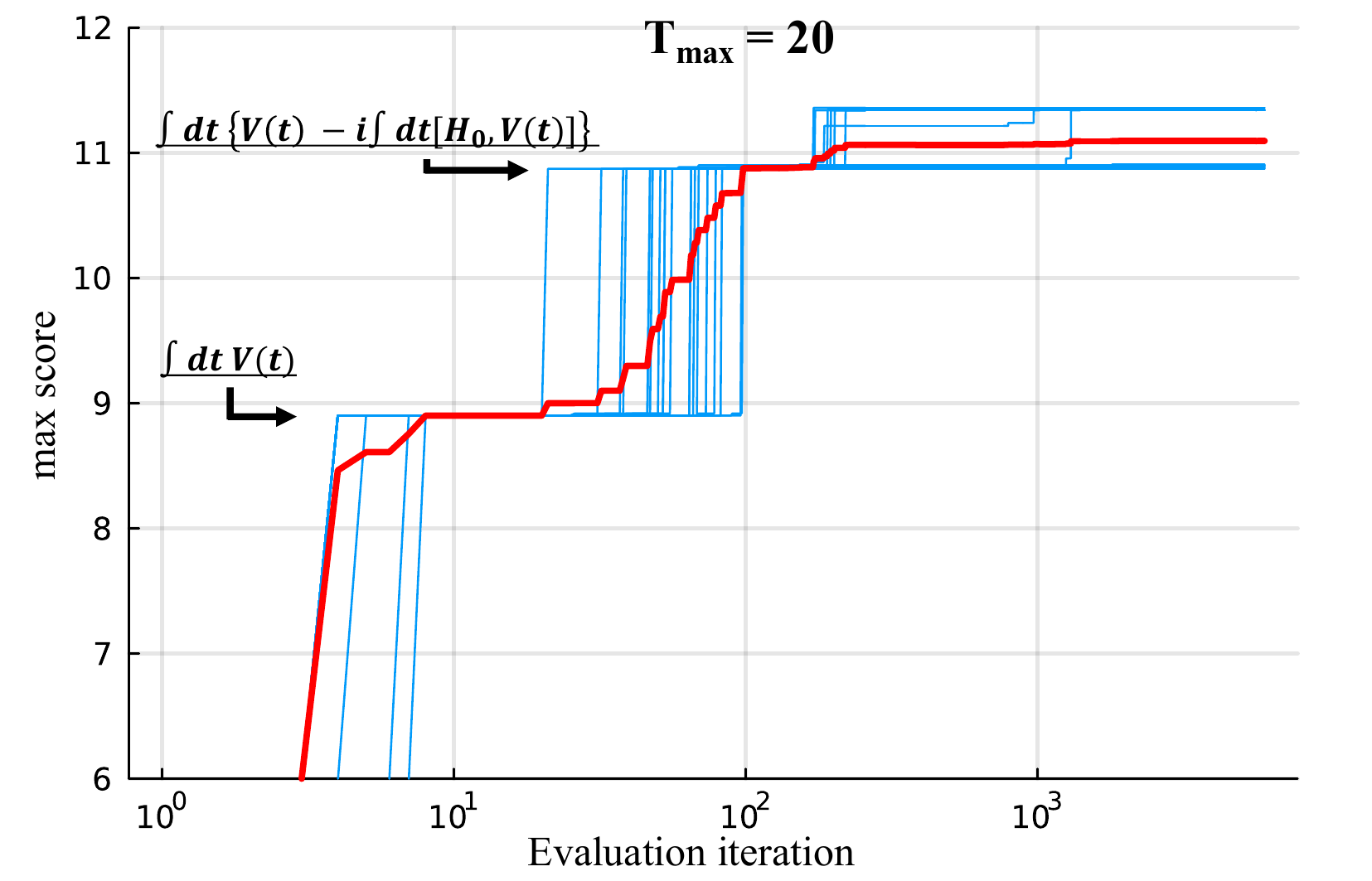}
    \includegraphics[width=0.48\linewidth]{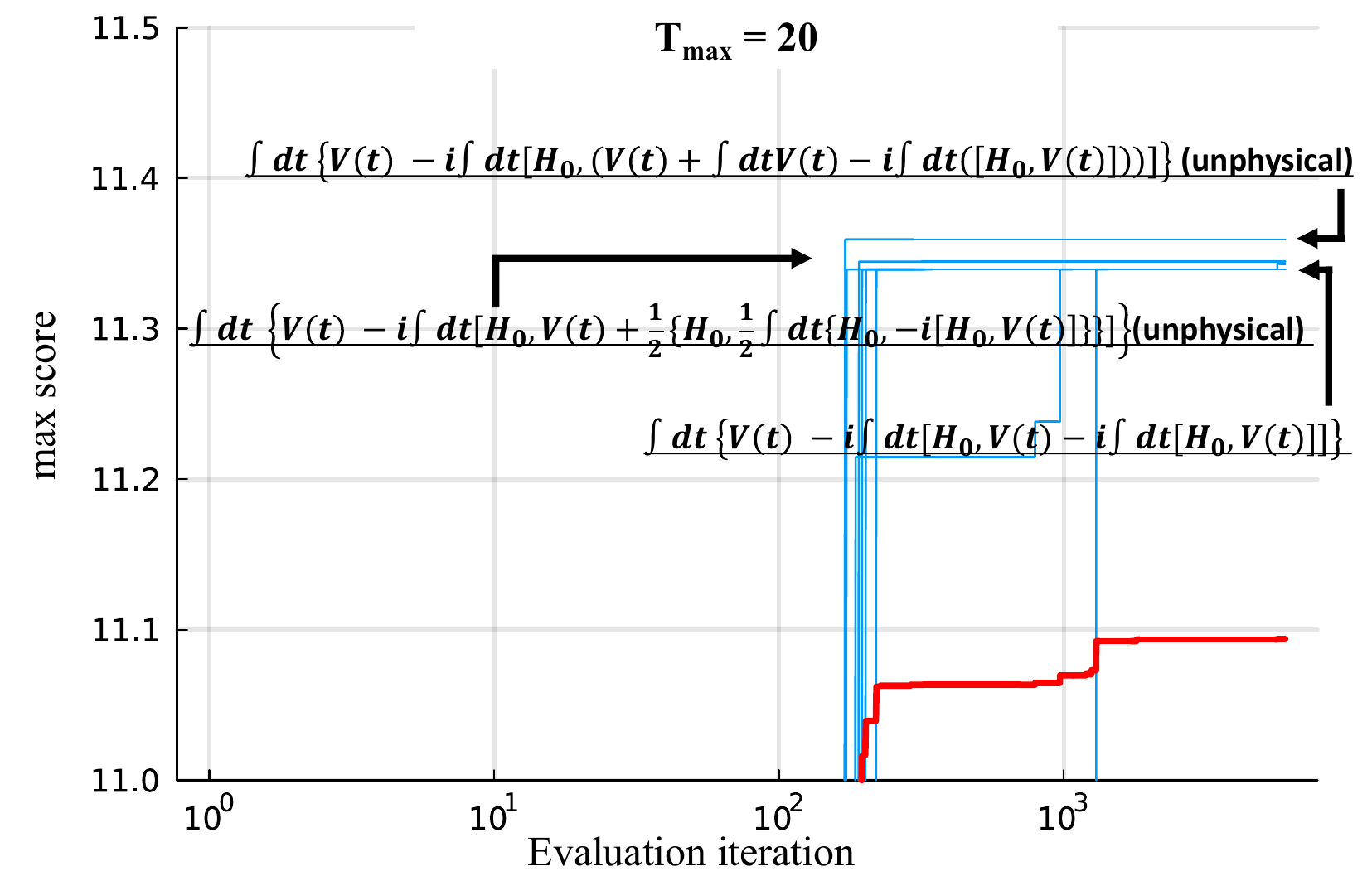}
    \caption{Searching dynamics by AZfP when we set $T_\mr{max}=20$. The y-axis represents the max score found in the search, and the x-axis represents the iteration number of calculated scores in the search. Because the search includes randomness, we have checked in 20 trials. The blue lines represent the performance of each trial, and the red line represents their average. Because the exploration space is wider than $T_\mr{max}=14$ case, the performance in the same evaluation iteration becomes worse. }
    \label{fig:long}
\end{figure*}



\end{document}